\documentclass{article}

\usepackage{arxiv}

\usepackage[utf8]{inputenc} 
\usepackage[T1]{fontenc}    
\usepackage{hyperref}       
\usepackage{url}            
\usepackage{booktabs}       
\usepackage{amsfonts}       
\usepackage{nicefrac}       
\usepackage{microtype}      
\usepackage{lipsum}
\usepackage{graphicx}

\title{The structure of behavioral data}

\author{
    Aurélien Defossez
   \\
    Department of Behavioral and Cognitive Sciences \\
    University of Luxembourg \\
  Esch-sur-Alzette, Luxembourg \\
  \texttt{\href{mailto:aurelien.defossez@uni.lu}{\nolinkurl{aurelien.defossez@uni.lu}}} \\
   \And
    Morteza Ansarinia
   \\
    Department of Behavioral and Cognitive Sciences \\
    University of Luxembourg \\
  Esch-sur-Alzette, Luxembourg \\
  \texttt{\href{mailto:morteza.ansarinia@uni.lu}{\nolinkurl{morteza.ansarinia@uni.lu}}} \\
   \And
    Brice Clocher
   \\
    Department of Behavioral and Cognitive Sciences \\
    University of Luxembourg \\
  Esch-sur-Alzette, Luxembourg \\
  \texttt{\href{mailto:brice.clocher@uni.lu}{\nolinkurl{brice.clocher@uni.lu}}} \\
   \And
    Emmanuel Schmück
   \\
    Department of Behavioral and Cognitive Sciences \\
    University of Luxembourg \\
  Esch-sur-Alzette, Luxembourg \\
  \texttt{\href{mailto:emmanuel.schmuck@uni.lu}{\nolinkurl{emmanuel.schmuck@uni.lu}}} \\
   \And
    Paul Schrater
   \\
    Department of Computer Science \\
    University of Minnesota \\
  Minnesota, IN 12345, USA \\
  \texttt{\href{mailto:schrater@umn.edu}{\nolinkurl{schrater@umn.edu}}} \\
   \And
    Pedro Cardoso-Leite
    \thanks{corresponding author.}
   \\
    Department of Behavioral and Cognitive Sciences \\
    University of Luxembourg \\
  Esch-sur-Alzette, Luxembourg \\
  \texttt{\href{mailto:pedro.cardosoleite@uni.lu}{\nolinkurl{pedro.cardosoleite@uni.lu}}} \\
  }

\begin{document}
\maketitle

\def\tightlist{}

\begin{abstract}
For more than a century, scientists have been collecting behavioral
data---an increasing fraction of which is now being publicly shared so
other researchers can reuse them to replicate, integrate or extend past
results. Although behavioral data is fundamental to many scientific
fields, there is currently no widely adopted standard for formatting,
naming, organizing, describing or sharing such data. This lack of
standardization is a major bottleneck for scientific progress. Not only
does it prevent the effective reuse of data, it also affects how
behavioral data in general are processed, as non-standard data calls for
custom-made data analysis code and prevents the development of efficient
tools. To address this problem, we develop the Behaverse Data Model
(BDM), a standard for structuring behavioral data. Here we focus on
major concepts in behavioral data, leaving further details and
developments to the project's website
(\url{https://behaverse.github.io/data-model/}).
\end{abstract}


\hypertarget{introduction}{%
\section{Introduction}\label{introduction}}

Experimental psychologists have been collecting behavioral data for over
a century now. As psychological sciences and related fields are
maturing, it has become increasingly clear that the field needs to
establish and converge on standards and standard operating procedures.

Data is essential to science. The recent rise of the open science
movement and the increased propensity to share and reuse data, as well
as the need to integrate results across multiple studies (e.g., within
meta-analyses) has revealed many shortcomings in the way we currently
process our datasets and has motivated several initiatives aiming to
make these datasets easier to find and use. Prominent examples include
BIDS (Brain Imaging Data Structure, \url{https://bids.neuroimaging.io/}
; see Gorgolewski et al. \protect\hyperlink{ref-gorgolewski2016}{2016}),
which focuses on brain imaging data and NeuroData without border
(Teeters et al. \protect\hyperlink{ref-teeters2015}{2015}) which tackles
neurophysiological data.

Behavioral data, however, has received comparatively less attention,
perhaps because at glance sight it appears simpler than those large
imaging datasets. We argue that behavioral data is in fact more complex
than meets the eye and that defining clear standards for behavioral data
may benefit all fields that rely on such data.

Standardizing how we define, name, format, organize, describe and store
behavioral data can provide multiple benefits, including:

\begin{itemize}
\tightlist
\item
  efficiency (e.g., less work, reuse of code, automated software);
\item
  robustness (e.g., less errors because of ambiguous idiosyncrasies);
\item
  transparency (e.g., fewer hidden choices in the code and data);
\item
  quality (e.g., via automated checks of data quality, consistency and
  completeness);
\item
  usability (e.g., via clear documentation, ready-to-use data).
\end{itemize}

Note also that non-standardized data formats call for non-standardized
data analyses which may obfuscate results at a time where more papers
are published than anyone can read. By contributing and using data
standards, we may accelerate scientific progress in psychological
sciences, as seems to have been the case in other fields (for examples,
see Teeters et al. \protect\hyperlink{ref-teeters2015}{2015}).

Here we present key ideas, concepts and principles that guided us in
creating the Behaverse data model (v2020.12.1); the more detailed,
somewhat opinionated and continuously updated specification of this data
model is accessible at \url{https://behaverse.github.io/data-model/}.
While there have been significant efforts to make behavioral data easier
to share and find, our focus here is on structuring behavioral datasets
to both reveal the essential structure common to behavioral data and
make them easier to (re)use.

\hypertarget{challenges-of-behavioral-data}{%
\section{Challenges of behavioral
data}\label{challenges-of-behavioral-data}}

There are key challenges to systematizing behavioral data.

First, behavioral data is highly diverse, as it includes body movement,
gaze, key presses, mouse clicks, written output and speech to name just
a few. We currently have no clear standards for each of these
measurement types, no standards that would be consistent across
measurement types and no standards on how to relate multiple measurement
types (both conceptually and practically). Hence, while we are
technically able to record rich, multivariate behavioral datasets, we
lack the conceptual and software tools to effectively exploit that
richness.

Second, to interpret behavioral data it is necessary not only to
characterize the behavior itself but also the context in which that
behavior occurred. Taking as an example the most basic of cognitive
tests, a particular key press is interpreted as being a response to a
particular stimulus within a particular task that evaluates to
``correct'' or ``incorrect''---the key press on its own, however, is not
very informative. Note that this is not necessarily the case for other
types of measurements (e.g., functional connectivity between two brain
areas). Hence, the accurate description and effective processing of
behavioral data requires rich annotations of the task and its underlying
theoretical constructs, the stimulus and the person's state. Major
efforts have been made in this direction (e.g., Poldrack et al.
\protect\hyperlink{ref-poldrack2011}{2011}); however, current solutions
haven't yet matured enough to be an integral and standard part of the
behavioral data analysis process.

Third, and related to the previous point, the way we describe behavioral
data is limited by our understanding of what a task is. Indeed, although
``tasks'' or ``tests'' are the cornerstones of experimental psychology
and related fields, we do not have a theory of tasks (which could for
instance characterize the structural relationships between any two
tasks) or even a clear framework on how to name or think about
fundamental concepts like ``instructions'', ``feedback'' or ``trial'',
let alone how to convert them into usable data structures---this applies
not only to concepts in psychology but also more general concepts like
``raw data''. This lack of clarity on concepts that are pervasive in
behavioral data have led to the discarding of what seems to us to be
critical information (e.g., task instructions not being recorded
anywhere) and is at least partially responsible for the large
inconsistencies one may find today across publicly shared datasets
(e.g., names, meanings and units of measurement). Hence, there is a
clear need to better conceptualize tasks, clarify concepts and converge
on standards.

Finally, the current practices and software tools used today for
behavioral data analyses seem inadequate to handle the rich and complex
data structures that seem necessary to accurately describe behavior.
Without a clear understanding of those data structures we can't create
effective tools that exploit that richness; but without effective tools
there is no incentive for researchers to invest effort in structuring
their data accordingly. Hence, until we have clear standards,
well-structured rich datasets, effective data analysis software and a
demonstration of added value, most researchers will understandably
continue to work the way they've done in the past. Hence, while we
should aim for better standards and tools, we still need to take into
account current practices and tools and offer solutions that can be
useful today.

The challenges we just described are considerable and overcoming them
will require sustained efforts over many years. Our goal here is to
contribute to overcoming these challenges and improve the way we
describe and organize behavioral data. The solutions we propose here
focus on three dimensions:

\begin{itemize}
\tightlist
\item
  \textbf{clarity.} Below we describe various ways in which current
  datasets are inconsistent. We then present and define several key
  concepts for behavioral data, the most important of which being
  perhaps the notion of a ``trial'' which we define as an instance of a
  ``task-pattern''. Rows in a ``trial table'' are then formed by
  extracting data from event data according to a task-pattern (using a
  query-like process) and each row in the ``trial table'' needs to
  contain all the information that is necessary to evaluate that trial
  (i.e., determine whether the response was correct or not). We also
  define different types of data tables (e.g., ``L1'' data) as well as
  canonical data tables (see
  \url{https://behaverse.github.io/data-model/}).
\item
  \textbf{consistency.} There are many choices to make when structuring
  data. These include, for instance, which naming conventions to adopt
  (e.g., ``RT'' versus ``response\_time''), which specific names to use
  for a particular concept (e.g., ``subjects'' versus ``participants'')
  and in what units to express certain variables (e.g., ``seconds''
  versus ``milliseconds''). While many of these choices may be
  arbitrary, it is vital for achieving the overarching goal of
  consistency to actually make these choices and document them in a
  clear way (Martin \protect\hyperlink{ref-martin2009}{2009})---we have
  started this process and documented our choices publicly (see
  \url{https://behaverse.github.io/data-model/}).
\item
  \textbf{usability.} Our particular choices for structuring behavioral
  data is motivated by the desire to make this data model useful and
  compatible with the tools and processes most researchers already use
  today. More specifically, we focus on tabular data (rather than more
  complex data structures) and aim for a good balance between human
  readability and computer/data efficiency. As we describe below,
  behavioral data involves many different types of data which could be
  compactly stored in a wide range of related tables. Such tables would
  however be much harder to process for humans as the information about
  a particular trial would now be distributed over multiple tables.
  Instead, we define, a primary ``trial table'' that contains all of the
  high level information about a trial (in line with current practices),
  and whose primary key serves to connect additional, possibly subtrial
  data (e.g., the timestamp of each of the images presented during that
  trial).
\end{itemize}

To keep this paper short, we focus here only on what we believe to be
central ideas; more content and specifics are available in the
accompanying website (\url{https://behaverse.github.io/data-model/}).

\hypertarget{data-consistency-levels}{%
\section{Data consistency levels}\label{data-consistency-levels}}

In this section we describe how typical behavioral data currently
available in public repositories look like and detail various issues
that make it hard to reuse them. Behavioral data from experiments in
psychology or related fields are currently scattered across multiple
locations, including researchers' personal webpages or various public
repositories (e.g., \url{https://osf.io})---which over the past decade
have made it much easier to find relevant datasets. Exploring these
datasets quickly reveals large differences in how behavioral datasets
are formatted, named, organized, described and shared---sometimes even
within the same lab. Unfortunately, finding a behavioral dataset today
is no guarantee that it will be usable at all and it seems that in most
cases substantial work would be necessary to understand and use them.

\begin{table}
\caption{Data Consistency Levels. It is our understanding that current standards in behavioral sciences places us within levels 0 to 1.}
\centering
\begin{tabular}{rl}
\toprule
Level     & Description\\
\midrule
  0 & The dataset is incomplete; critical information is missing (e.g., description of what the variables mean).\\ 
  1 & All datasets are formatted in a unique way and can't be joined without reformatting.\\
  2 & Datasets can be joined when they originate from the same task "variant" (e.g., a 2-back task using digits) \\
    & but not from distinct variants (e.g., a 2-back versus a 3-back task).\\
  3 & Datasets can be joined across all variants of a a task (e.g., all N-back tasks).\\
  4 & Datasets can be joined within a family of tasks (e.g., all CPT-like tasks).\\
  5 & Datasets can be joined across several task families.\\
  6 & All datasets can be joined.\\
\bottomrule
\end{tabular}
\label{tab:DCScale}
\end{table}

To qualify the current state and future progress in behavioral data
standardization we devised a \emph{data consistency} scale which
describes 7 levels of consistency, defined by the type of table
joints---or merging of different data tables---that a data model
supports (see Table \ref{tab:DCScale}). Next, to get a rough sense of
the data consistency level in cognitive psychology, we selected three
popular cognitive tests---the digit-span task, the N-back task and the
AX-CPT task. We then searched, downloaded and reviewed recent datasets
from \url{https://osf.io}. Our goal here is not to make claims about the
quality of the specific data samples we chose or of the research
conducted using that data (hence, we keep them anonymous). Our goal is
also not to be exhaustive and have a definite characterization of the
current state of affairs. Instead, we want to point out the diversity
and inconsistencies that currently exist in such datasets and describe
the various issues that one encounters right after discovering what
seems to be a relevant dataset. Below we describe these issues in the
order one would encounter them.

\hypertarget{inconsistent-data-formats}{%
\subsection{Inconsistent data formats}\label{inconsistent-data-formats}}

Most data sets seem to be in csv format. However, we also found several
Excel files and proprietary formatted data which could not be read at
all. Oftentimes, data is shared as a single data file (containing the
data for all participants) or in multiple files that all have the same
structure (e.g., one file per participant). These datasets rarely
provide a codebook to explain the meaning and possible values in their
datasets and it would therefore be necessary to manually go over other
available materials (e.g., the corresponding research paper) to attempt
to uncover that information.

\hypertarget{unknown-or-inconsistent-data-levels}{%
\subsection{Unknown or inconsistent data
levels}\label{unknown-or-inconsistent-data-levels}}

Behavioral data come in various levels of granularity. Some data sets
might contain each response given by every participant while others may
only include aggregated data for each person (e.g., one row per
participant versus one row per trial). It is typically impossible to
know which level of data granularity the shared data offers before
actually opening and inspecting the data files.

It is also very common that data tables mix data that are from different
sources or levels of granularity. For example, a data table might
include trial-level data for each participant (i.e., a row for each
response the participant gave) but at the same time have a column that
indicates the age and gender of the participants (e.g., the values
``21'' and ``female'' repeated across all rows within a given
participant) or even summary statistics (e.g., d'prime), whereby it can
sometimes be ambiguous as to whether those summary statistics were
computed on the trial-level and then joined to the trial-level data or
whether they were computed using other data.

\hypertarget{inconsistent-variable-naming-conventions}{%
\subsection{Inconsistent variable naming
conventions}\label{inconsistent-variable-naming-conventions}}

Naming variables is notoriously hard and unsurprisingly, there are
numerous inconsistencies in variable names (Martin
\protect\hyperlink{ref-martin2009}{2009}). We found inconsistencies in
naming conventions across but also within datasets. Some data sets use
lower-case ``snake\_case'' (e.g., ``n\_correct'') others use upper-case
snake-case (e.g., ``N\_Level''). Some use CamelCase (e.g.,
``TrialList'') or a mixture between CamelCase and snake\_case (e.g.,
``V\_FalseAlarm'') or still something else (e.g., ``TrialList.Sample``).
Some variables may be in all uppercase (e.g., ``CUE\_ACC'') or include
information about the coding scheme (e.g., a column named ``FEMALE=1'').
While one may argue that such conventions are more or less arbitrary, it
stands to reason that a given convention should be used consistently
across a given dataset. This is not the case in the random sample of
studies we've reviewed as within the same table we could find for
example ``Span\_amount'', ``CorrectAnswer'' and ``TrialList.Sample``.

We also note the variability with which the same construct is named and
coded. For example, most if not all datasets have a variable to refer to
individual participants in a study. Common variable names to refer to
participants are ``id'', ``Subject'' and ``SubjectID''. The use of
``id'' may however be ambiguous (id could perhaps refer to trial index).
Sometimes the values that this variable takes is an integer (e.g., 15),
sometimes it's a concatenation of something that seems to be a study or
condition name and an integer (e.g., ``A\_15''). Coding schemes for the
subject variable may be somewhat arbitrary but there might be an issue
when there are multiple datasets. For example are ``A\_15'' and
``B\_15'' different people or are they the same person (participant 15)
that completed two different tasks (``A'' and ``B'')?

Another variable that is common in behavioral data sets refers to
individual trials within an experiment. Again we observed quite some
variability. While it is common to use the name ``trial'' or ``id'', we
also found datasets where the trial index variable was missing and
seemed thus to be implicit in the order of the rows of the table and
other cases where the ``trial'' variable was not used to refer to the
index of the trials but rather to describe a type of trial (e.g.,
``start'', ``nontarget'', ``v\_target'').

\hypertarget{unknown-values-and-units}{%
\subsection{Unknown values and units}\label{unknown-values-and-units}}

Another common issue, which might be resolved by the use of codebooks,
is the absence of information about the possible values a variable can
take and what units a variable is expressed in. For example, it is very
common for data sets in experimental psychology to include response time
data. It is typically not possible to determine if they are expressed in
milliseconds, seconds or minutes before inspecting the data and using
domain knowledge to infer the units.

\hypertarget{conclusion}{%
\subsection{Conclusion}\label{conclusion}}

A quick review of publicly available datasets reveals substantial
inconsistencies in the way individual researchers/research groups
(including ourselves) structure their data. Such inconsistencies are
inconsequential for researchers working on their own data but limit the
reuse of data by other researchers and the aggregation across data sets,
even for datasets collected using very similar tasks.

In what follows we first describe some key properties of behavioral data
before introducing the behaverse data model we currently use.

\hypertarget{behavioral-experiments-require-multiple-types-of-data}{%
\section{Behavioral experiments require multiple types of
data}\label{behavioral-experiments-require-multiple-types-of-data}}

Data from cognitive psychology experiments are often shared in the form
of a single table where each row refers to an individual trial completed
by a person. While it is convenient to only have one file for
data-analysis, this ``simplicity'' is in fact illusory and valuable data
is currently hidden within the associated paper, computer code (or still
other documents), if not missing altogether.

Typical behavioral data collection scenarios involve collecting data
that are semantically distinct but intrinsically linked by virtue of the
data collection situation. Consider for instance a typical cognitive
psychology experiment. A research group invites participants to their
lab to complete a computerized version of the ``digit-span'' test twice.
What type of information could one expect this study to collect? Below
is a non-exhaustive list of the kinds of data that are or should be
recorded:

\begin{enumerate}
\def\labelenumi{\arabic{enumi}.}
\tightlist
\item
  information about the study (e.g., who conducted the study, when and
  where; what was the intentions; is the study approved by an ethics
  committee; what was the funding source); this information is typically
  idiosyncratically present in manuscripts but should be structured in a
  standard way, for example, in a ``Study'' table.
\item
  information about the participants. This can include variables like
  birth date, gender, or nationality. Part of this information may be in
  the manuscript (e.g., ``we recruited participants from city X'') and
  part of it may be in the trial data (e.g., the ``age'' and ``gender''
  variables that are in the trial-level data). It is important to note
  that some information about participants is fixed (e.g., birth date)
  while other information may be context dependent and linked to the
  actual moment of data collection (e.g., age). Static information about
  the participant should be stored in a ``Subject'' table, while
  dynamically changing information (e.g., age) might be stored in a
  ``Session'' table.
\item
  information about the activity participants engaged with. In cognitive
  tests, this would include for instance the name of the task, task
  parameters, the instructions given to participants. This information
  is typically buried in a research paper and often incomplete (e.g.,
  the actual task instructions, although essential, are rarely listed in
  full). More and more often, the actual code that was used to run the
  activity is made available as well---but it may require significant
  work to uncover task parameters from code. Information about the task
  or activity should be organized in an ``Activity'' table.
\item
  Information about the hardware being used and of participants'
  physical environments. For example, this could indicate particular
  brands and models of tablets or computers, versions of OS and
  software.
\item
  Information related to the interactions between the participant and
  the computer/environment, in particular information about what
  stimulus was shown, when and where and what inputs participants made.
\item
  Information about events that occurred while participants were engaged
  in the activity. For example, this could include information about the
  quality of the data collection process (e.g., average frame rate) or
  observations made during the experiment (e.g., experimenter notes that
  a participant seems to be falling asleep); this type of information
  might be stored in a lab or personal notebook.
\item
  Information about participants progress through the study (e.g., list
  of participants having completed one test but not the other, data and
  time of completion of tasks, order of task completion).
\end{enumerate}

The list above is not exhaustive but includes the main types of data
that could in principle be collected in all behavioral experiments. The
point we want to make here is that a data collection campaign comprises
in fact multiple data tables and each data table has its own type (i.e.,
specific requirements, formats).

Our goal in this document is not to go over each of these data types and
review existing solutions (although such an enterprise would certainly
be useful). Our primary focus in this document is on the data type (5)
which we'll refer to as the actual \textbf{behavioral} data. In our
opinion, this is the data type that has received the least attention and
presents the largest inconsistencies across studies. It is also the type
of data that is most relevant for behavioral data analysis and which
would most benefit from standardization.

\hypertarget{behavioral-interaction-data}{%
\section{Behavioral, interaction
data}\label{behavioral-interaction-data}}

There is a lack of clarity on the meaning of terms that are commonly
used in behavioral data (e.g., what constitutes ``raw data''? what is a
``trial''? what is a ``task''). In
\url{http://behaverse.org/data_model/} we define several of those terms
and other conventions we use in the behaverse data model. In what
follows, we attempt to present the big picture view of behavioral data
and clarify essential terms.

\begin{figure}[ht]

{\centering \includegraphics[width=1\linewidth]{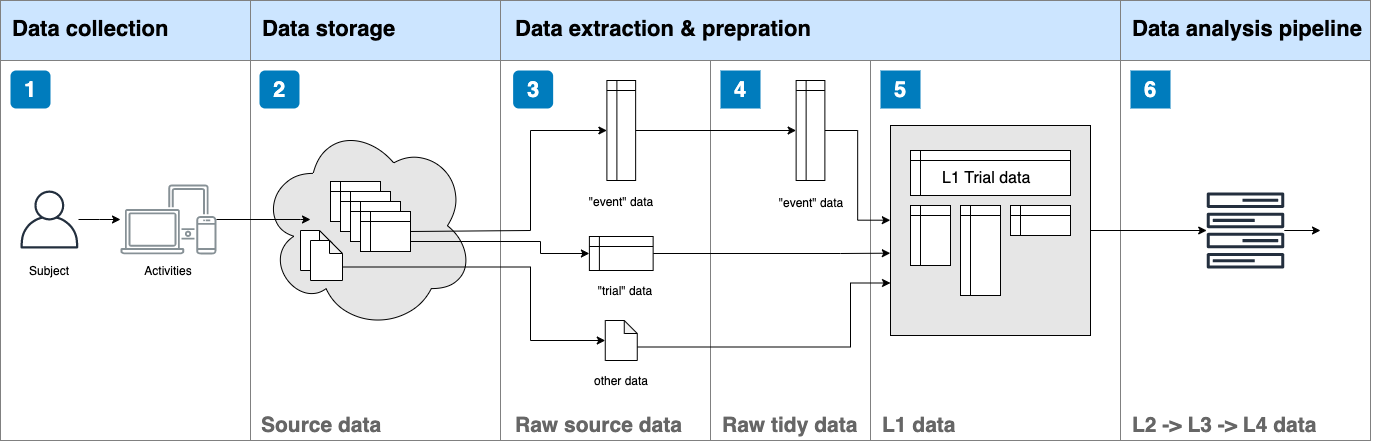} 

}

\caption{From data collection to analysis. 1) Subjects interact with digital artefacts and produce data. 2) The resulting data (“source data”) is typically stored in idiosyncratic formats, possibly determined by technical constraints of the digital artefacts. Furthermore, this “source data” may contain data that is not of direct relevance to researchers (e.g., technical information about the software) and important information may come from other sources (e.g., information about the study that is present only in the corresponding research paper). 3) It is typically necessary to extract the relevant data from the source data. Here we distinguish “event” data and “trial” data. Event data describes the behavioral data as a sequence of time stamped events, which have specific types (e.g., a mouse click) and data (e.g., the screen coordinates of the click). Trial data organizes those events following a task-pattern into a tabular form, where each row describes  one trial. Further data files are necessary for example to describe the study. Note that it is typical for the data collection artifacts to already embed some data processing code and keep as source data only the “trial” data. 4) The most important type of behavioral data appears to be the event data from which different trial datasets may be extracted—this is in our opinion what should be viewed as the raw data and it will be valuable in the future to standardize behavioral event data and develop effective tools to deal with such data and extract trial-based data from them. 5) We define as Level 1 data, the data tables which are organized by trial. These are the tables we believe are most useful given current practices. In particular, we define the L1-Trial table, where each row contains complete and standardized information describing a particular trial (as is already currently the case, albeit inconsistently) and where the trial identifier is used as a primary key to additional, more detailed or specific tables (e.g., a table describing each of the mouse clicks that occurred during a trial). 6) The L1 data serves as the standardized input to data processing pipelines, which will derive additional tables (e.g., L2, L3), for example by transforming and summarizing data or aggregating across subjects.}\label{fig:pipeline}
\end{figure}

\hypertarget{source-data-raw-data-and-derived-data}{%
\subsection{Source data, raw data and derived
data}\label{source-data-raw-data-and-derived-data}}

We consider as source data, all the data that is saved by the data
collection artifact (e.g., computerized cognitive test) in its original
structure and format (e.g., a single data file in a proprietary data
format; multiple json files). Source data can contain all sorts of data.
It includes the raw data but may also include metadata (e.g.,
information about the artifact itself) as well as derived data (e.g., a
performance score computed from the raw data). Source data is typically
in idiosyncratic formats and not usable as is.

Not all source data is raw data; and raw data needs not be source data.
There are certain operations that can be performed on the raw source
data to extract and constitute a dataset that is more usable without
that dataset losing the ``raw data'' status. For example, if a source
file is saved as a csv (comma separated values) file, converting that
csv file into a tsv (tab separated values) file, is a trivial operation
that has no consequences on the outcome of the study. On the other hand,
filtering out some data based on performance or rounding numeric values
are operations that may impact the outcome of subsequent analyses; hence
the data that results from applying those operations can no longer be
considered ``raw''.

Operations we consider to preserve ``rawness'' are selection by type
(not by value), removal of duplicates, renaming of variable names for
clarification, change of units, reordering of rows and columns and
referencing/indexing (e.g., numbering rows of a certain type) and
reversible file format conversion (e.g, csv to tsv). In short, as long
as the information in the data is equivalent to the information in the
raw source data, in our opinion, that data can be said to be raw.

\hypertarget{event-data-and-trial-data}{%
\subsection{Event data and trial data}\label{event-data-and-trial-data}}

Two common ways to structure behavioral data are by event or by trial
(source data may contain either event data or trial data or both). Event
data lists particular events that occurred during a study (e.g., a
person pressed a key, a stimulus was displayed on the screen) with a
timestamp (i.e., when did that event occur) and information describing
the event (e.g., where on the screen did the click occur, how long did
it last). The event data format is common in cases where behavior is
related to other, time varying measures (e.g., in fMRI or EEG studies);
it is much less common in behavioral sciences where information about
when particular events occured is often discarded. In those fields, it
is much more common to structure the behavioral data by trial, meaning,
as a table where each row corresponds to a ``trial'' and each column to
a variable describing what happened during that trial (e.g., for
trial\_index = 3, correct = TRUE).

It is important to note that beyond the shape factor, trial data and
event data are quite different. Event data may describe events as they
occurred and are thus more objective (e.g.~a click occurred at timestamp
6.824). Trial data, on the other hand, are fundamentally tainted by the
experimenter who needs to define (typically implicitly) a
``task-pattern'' which defines which events to select from the flow of
events that occurred during the study and how to aggregate and/or
transform them in order to constitute a row in the Trial table.

Let's take an example to make this point clearer. In a N-back task,
participants are shown letters, one at a time, and asked to report
whether the letter that is currently displayed is the same as the letter
shown N steps earlier. Let's further compare a 2-back and a 3-back test
that use the exact same sequence of letters. The event data from these
two tasks may look virtually identical (they have events describing the
occurence of letters and key presses). The trial data, on the other hand
should look differently because for the 2-back test we use a different
``task-pattern'' than in the 3-back test. For example, in the first case
we might describe the stimulus of the first two trials as ``3-1-3'' and
``1-3-4'', while the same sequence of events in the 3-back task only
forms one trial whose stimulus could be described as ``3-1-3-4''.

Figure \ref{fig:pipeline} shows various steps in the lifetime of a
dataset, ranging from its collection to the aggregation of summary
statistics across participants. The format and structure of the source
data is subject to various engineering constraints and specific to
particular data collection software systems; it is therefore unlikely
that we'll converge on standards for source data that would apply to all
use-cases any time soon. However, we could aim to define standards for
raw event and trial data which could be readily used as input for data
analyses pipelines and shared on public data repositories.

Here we focus on describing the L1 data, leaving for later
standardization efforts of event data. This choice is motivated by our
belief that standardizing trial data will be of most practical value to
the research community.

\hypertarget{key-concepts-for-specifying-trial-data}{%
\subsection{Key concepts for specifying trial
data}\label{key-concepts-for-specifying-trial-data}}

The data format that seems most useful and characterizes many shared
behavioral datasets displays one row per ``trial''---we call this the
``Trial table''. For example if an experiment tested 50 participants and
each participant completed 200 trials, the Trial data table would
contain 10'000 rows in total (assuming all the data was in a single
table).

It is important to note at this stage that the term ``trial'' is not
used in a consistent manner in the literature and the corresponding data
files. The following section aims to highlight and clarify this issue.

\hypertarget{the-meaning-of-trial}{%
\subsubsection{The meaning of ``trial''}\label{the-meaning-of-trial}}

Different meanings are associated with ``trial''. Firstly, ``trial'' may
be used to refer to iterations of a chunk of code that is executed
repeatedly (or equivalently a sequence of stimulation and input
recording events). For example, a trial may consist of the presentation
of an image on the screen and the recording of a keypress made by the
user after the appearance of that visual stimulus. Secondly, ``trial''
may be used as an index to refer to individual rows in a data table. For
example, each time the user presses a key we add a line to a data table
that indicates which stimulus was shown and which button the user
pressed. Thirdly, ``trial'' may refer to an instance or sample of a
specific experiment in the statistical sense. For example, we want to
determine if a particular coin is biased and repeatedly throw that coin
and record the outcome; each throw represents a trial of that particular
experiment. Finally, ``trial'' may be used to refer to a period of time
or ``episode'' during the experiment (e.g., ``the participant blinked
during the second trial'', ``there was a 5 minutes break between trials
50 and 51''). In the most basic cognitive tests, all three meanings are
congruent and thus interchangeable. But as experimental designs increase
in complexity, even slightly, those notions are no longer equivalent and
it becomes necessary to use more precise terminology.

Let's take a simple example to illustrate this point. Imagine a task
where a letter is shown for 1 second and participants have to press one
of two keys in response to that letter during the subsequent
second---this code loop then repeats 100 times. In condition-1,
participants are asked to press the right key each time they see the
letter X and to press the left key otherwise (a ``Sustained Attention to
Response Task'' like test Robertson et al.
(\protect\hyperlink{ref-robertson1997}{1997})). In condition-2, users
are asked to press the right key each time they see the letter X but
only if it was preceded by the letter A and to press the left key
otherwise (the AX-CPT task; Braver et al.
(\protect\hyperlink{ref-braver2001}{2001})). Finally, in condition-3,
both tasks are to be completed at the same time: a single letter is
successively shown on the screen, but there are now two sets of buttons,
one per task.

While the same code can be used to run these three conditions, from the
perspectives of the participant and researcher, they are different in
important ways. In condition-1, we would expect the stimulus description
to refer to a unique letter, while in condition-2, a stimulus would
refer to pairs of letters (this information is necessary to determine in
each case whether participants' responses were correct or not).
Furthermore, if condition-1 and condition-2 use the same sequence of
letters, the resulting number of trials will be different across the two
conditions. Consequently, in this example, a ``trial'' in the code-loop
sense no longer maps directly to a ``trial'' in the table index sense as
information from two different code-loop trials is now contained in a
single table-index trial. Next, if we consider the second experimental
condition, one might assume that an experimenter will be interested only
in those instances where a letter X was shown and it was preceded by
another letter. If those instances define ``trials'' in the statistical
sense, then trials should count only these specific instances. For
example, if we assume that there were 100 code-loop trials (i.e.,
presentations of letters) but only 5 of those presented the letter X
then there could at most be 5 trials (in the statistical sense) in that
experiment, and thus only 5 rows in the corresponding data table.
Finally, if we focus on condition-3, we see that for a given letter,
there are two ``trials'' (one per task) occurring at the same time.
Trial in this (and other cases) can therefore no longer be used to refer
to a time period---to refer to particular, temporally distinct and
non-overlapping time periods in an experiment we recommend to use
``episode'' instead. In condition-3, we could then have the same episode
index correspond both to the 5th trial of the first task and the first
trial of the second task.

The example above illustrates that ``trial'' can be used in inconsistent
ways and that it is necessary to clarify its meaning. Within the
behaverse data model we use the statistical definition of trial and
define a trial with a corresponding task-pattern (see below). For
indexing rows in a table we use a more generic ``id'' variable and for
indexing particular time periods in a study we use ``episode''.

\hypertarget{the-task-pattern}{%
\subsubsection{The task-pattern}\label{the-task-pattern}}

Consider again the example experiment presented earlier where under two
different conditions, letters were presented successively and
participants were required to press one of two keys in response to those
letters. The event data from both of these conditions could virtually be
identical, with the same type of events being recorded each time a
stimulus is shown or key is pressed. However, the corresponding trial
data would look rather differently across both sets of conditions.

One can think of the trial data as something that is ``created'' from
the event data (+ some other stuff). Indeed, one could write
``extraction'' code that would parse the event data looking for specific
sequences of event types, extract the data corresponding to those event
types and process and shape them into a row of the trial table---we call
this code the ``extractor'' and save its parameters together with its
trial data.

The specific sequence of event types, used by the extractor to query the
event data, is what we call the task-pattern (in analogy to pattern in
regular expressions). A task-pattern is typically of the form
\{stimulus-set; action-set\}. In condition-1 of our example task, the
stimulus-set might be all letters, while in condition-3 it might be all
pairs of successively presented letters or all pairs of letters where
the second letter is the letter ``X'' (depending on the experimenter's
intention). In both cases, the action-set is any of the two possible
button clicks that occur within 1 second after the stimulus.
Task-patterns can of course be more complex; the key idea here is that
the definition of a trial of a particular type is determined by a
task-pattern. In the behaverse data model, when we index a trial, we
index trials for a given task-pattern.

There are two points we want to emphasize here. Firstly, while the event
data can be seen as an objective description of what actually happened
during a study (e.g., the letter ``A'' shown on the screen center at
10:42:01''631''; the left arrow key was pressed at 10:42 02'246''), the
trial data necessarily reflects the experimenters view of what that data
means (e.g., the key press is a response to the letter, the response
time is computed as the difference of times stamps and equals 0.615
seconds, and the response is correct given the current task rule). In
fact, a different trial dataset could be generated from the same event
dataset. The take-home message then, is that a) we need to store the
event data as this data is privileged and more objective/raw than the
trial data, and b) for a given trial dataset we need to maintain
information about its provenance (e.g., the name of the task-pattern or
extractor-code used to go from event data to trial data). Secondly, we
believe that the concept of task-pattern is important beyond the context
of data extraction and might be useful to characterize tasks for
computational modeling or to implement artificial agents capable of
performing tasks.

\hypertarget{evaluation}{%
\subsubsection{Evaluation}\label{evaluation}}

The task-pattern defines what constitutes a valid trial within a given
experiment; it defines a subset of all possible stimulus and input
sequences. Each element in this set of valid trials is mapped to a
value. For example, it is very common in cognitive psychology for the
response on a given trial to evaluate to ``correct'' or ``incorrect''.
The value function or ``evaluation'' can be seen as a set of rules which
are typically (implicitly) described in the task instructions (e.g.,
{[}to be correct:{]} ``if you see the letter X press this key, otherwise
press that key''); the value function may also be defined relative to an
idealized policy---the particular way the experimenter believes
participants should map stimuli (sequences) to action (sequences) within
the context of the study.

\hypertarget{runtime-extraction-and-evaluation}{%
\subsubsection{Runtime extraction and
evaluation}\label{runtime-extraction-and-evaluation}}

It is important to note that the software we use to present stimuli to
participants and record their actions typically encodes information that
reveals our intentions and may in fact distort the data. For instance,
some researchers might not record event data and instead create the
trial data directly as events unfold in time---their code instantiates
an ``extractor''. This will typically discard data (e.g., when did a
trial start) which makes it impossible to later reconstruct the time
course of events as they occurred. Furthermore, that same code also
typically includes evaluation code, as this might be necessary within
the experiment itself, for example to display participants a
correct/incorrect feedback signal for a given response.

It can be convenient and sometimes necessary to have these data
processing functions embedded in the data collection code and operate
during runtime on the events as they occur. However, one should also be
wary of the fact that this code may contain errors. If we record only
the output of those processes, i.e., runtime generated trial data but no
event data, it might be impossible to detect and ultimately correct
those errors.

\hypertarget{trial-data-versus-l1-data}{%
\subsubsection{Trial data versus
L1-data}\label{trial-data-versus-l1-data}}

When describing the data that is extracted from the event data we used
both the terms L1-data and Trial data in the sections above. These two
terms, however, are not synonymous. Rather, L1-data refers to the state
of the data (typically multiple tables) within a stage of the data
analysis pipeline (see Figure \ref{fig:pipeline}). Trial-data, on the
other hand refers to a specific type of data table where each row
contains data from a single trial as defined above. In the next section
we'll review the structure of the L1-data, and discuss what other tables
besides the Trial table may exist within L1.

\hypertarget{l1-data-model}{%
\section{L1 data model}\label{l1-data-model}}

Behavioral data (e.g., from computerized cognitive tests) are typically
shared in a tabular format (e.g., one csv file per task), where rows
typically correspond to individual ``trials'' and columns refer to
different types of variables that describe that trial (e.g., response
time). This, however, is insufficient. Firstly, it is already the case
that the single-table trial-data does not include all necessary
information. For example, it is typically necessary to read the paper
about that data to learn about task parameters that did not vary across
trials (e.g., the duration of stimulus presentations). Extracting that
data and putting them in a consistent format would facilitate subsequent
data usage. Secondly, behavioral data contains information that can be
grouped into different semantic categories. These subcategories may have
nested structures which do not play well with a simple single-table
format but may instead be properly organized into multiple sets of tidy
tables. More specifically, we define the following semantic data
categories for the L1 data:

\begin{enumerate}
\def\labelenumi{\arabic{enumi}.}
\tightlist
\item
  \textbf{Context:} provides context information for a particular trial,
  such as, identifiers for a study, a session, a participant and task.
\item
  \textbf{Task Information:} describes the tasks participants were
  exposed to (e.g., instructions, task parameters).
\item
  \textbf{Extraction Information:} describes how event data was
  converted into trials.
\item
  \textbf{Stimulus Information:} describes what stimuli were presented
  to participants.
\item
  \textbf{Options Information:} describes the different options
  participants had for responding on a given trial.
\item
  \textbf{Input information:} describes the actions participants made
  (e.g., a button click).
\item
  \textbf{Response Information:} describes the meaning of participants
  inputs within the context of the task (e.g., option ``match'').
\item
  \textbf{Evaluation:} describes the value associated with participants'
  responses (e.g., this response was correct); this value is not
  necessarily communicated back to the participants.
\item
  \textbf{Feedback Information:} describes if and how participants
  received explicit information about their response or performance
  (e.g., green check after a correct response); this data describes
  physical events shown to the participants. Note that one may have the
  case where a ``green check'' feedback is shown to participants after
  an incorrect response (i.e., evaluation and feedback are distinct
  constructs).
\item
  \textbf{Outcome Information:} describes the consequences of the
  participants' action in the test. For example, in a serial ordered
  search task, participants are asked to open boxes to search for a
  token. Opening a box has the outcome of revealing its content and
  changing the state of the world (e.g., it reveals an empty box). While
  an outcome may implicitly contain feedback information, it is not
  necessarily the case. On the other hand feedback is solely meant to
  convey participants information about their performance. Outcome and
  feedback and evaluation are distinct constructs. In our box opening
  example, a participant may correctly click on an empty box
  (evaluation), see a green check (feedback), and see that the box is in
  fact empty (outcome).
\item
  \textbf{Reward Information:} participants sometimes get a reward in
  tests; this could for example take the form of points, money or even
  food.
\item
  \textbf{Experimental Design Information:} provides additional,
  optional data or features that the experimenter believes will be
  useful to interpret participant's responses (e.g., tagging certain
  trials in the N-back task as being ``pre-lure'' or ``post-lure'' with
  the intention to contrast performance on these two types of trials).
\item
  \textbf{Hardware information:} provides information about the hardware
  that was used to collect the data (e.g., this keypress was collected
  from keyboard \#2).
\item
  \textbf{Technical Runtime information:} provides information about how
  well the trial was executed from a technical point of view (e.g., were
  there unexpected lags?).
\item
  \textbf{Information about additional data:} provides information about
  additional measures that might have been collected during the study
  (e.g., brain imaging data).
\end{enumerate}

Each of these categories could have its own table with additional tables
associated to them because there are typically different subtypes of
data for each of these (for example, there are different kinds of
possible stimuli and each kind of possible stimulus could have its own
table).

There are two points we want to make here. First, behavioral data, as we
hope to have demonstrated, is more complex than typically assumed; it
involves a myriad of interconnected data tables. Second, current
practices and data analysis tools do not address this complexity and
instead focus on an easier to handle subset of the data (i.e., only the
data that is strictly necessary for a particular analysis).

\begin{figure}[ht]

{\centering \includegraphics[width=0.65\linewidth]{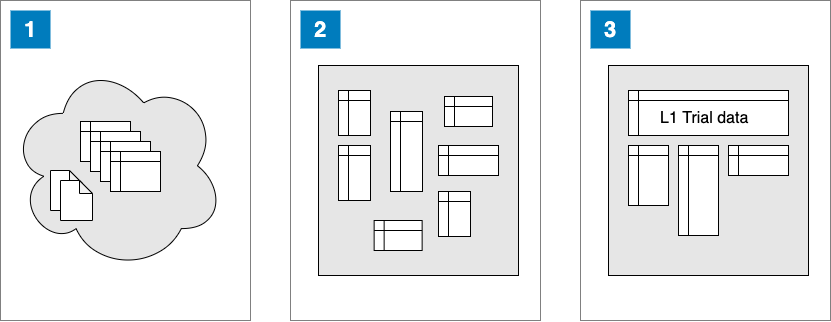} 

}

\caption{L1 Trial data. 1) In source data, relevant information may be scattered across multiple data files in a way that is not practical for subsequent processing. There are various design options to reorganize the source data into data structures that can be standardized and are easier to use. 2) One solution is to factor the data into many compact tables within a relational database system. While this solution has many technical advantages, it doesn’t play well with current practices. 3) An alternative design solution—the one we chose for the current behaverse data model— defines a main “L1 Trial” table which is similar to what researchers already use today. However, in addition to providing the trial data, the L1 dataset contains additional, related tables (as in 2). Tables in L1 are related to each other by various primary keys, the most important one being the trial identifier within the Trial table. We believe that this solution is both of practical use for researchers and offers the possibility to augment the Trial table in a principled way to capture more of the richness of behavioral data than is typically the case. }\label{fig:L1_diagram}
\end{figure}

In order to get a more comprehensive and consistent handle on all of the
behavioral data while at the same time remaining compatible with current
practices and tools we opted for a particular set of design principles
to organize the multiple L1 tables (see Figure \ref{fig:L1_diagram}).

The first principle is to keep a trial table which is similar to what is
already customary in the field. Each row in this table describes one
trial and columns may contain summary information about particular
aspects of that trial. For example, in a digit-span task where the
stimulus is a sequence of digits presented at a certain rate one may
summarise the stimulus for a given trial as ``3;4;5;1''. We define
standards and conventions for that trial table to achieve consistency
across datasets (see \url{https://behaverse.github.io/data-model/}).

The second principle is to separate information depending on whether or
not it is common or specific (e.g., to a task) and whether it describes
the trial as a whole or particular events that occurred during the
trial. Taking again the example of the digit-span test, ``3;4;5;1''
describes the stimulus at the trial level and is thus present in the
trial table. The timestamp of the digit 5 during that trial is specific
to an event and is thus present in the stimulus table which describes
all the stimuli that occurred within each trial.

The third principle is that the trial table serves as the master table
with the id of each row in that table serving as the key to link all the
tables within L1. For example, knowing from the Trial table that
``3;4;5;1'' was presented on trial\_id 2378, one can find within the
Stimulus table the list of stimuli shown during that trial together with
the properties of those stimuli (e.g., timestamp, location, duration).

We believe that this design strikes a good balance between the somewhat
contradictory requirements (e.g., the efficiency of a fully relational
database versus human readability and ease of use); it is compatible
with the way researchers are already structuring their trial data and
offers a principled way to organize related data that is currently
ignored but shouldn't.

\hypertarget{discussion}{%
\section{Discussion}\label{discussion}}

The standardization of behavioral data structures may not be the most
exciting endeavour for a researcher---after all, great scientific
advances were made without such standards, researchers can analyse data
without following standards and it may seem to many that time spent on
such mundane issues is time diverted from doing actual research. While
there certainly is some truth to those statements, we believe that
developing good standards for structuring behavioral data holds the
promise for significantly improving the quantity and quality of
behavioral research and may lead to novel insights.

As have argued many before us (e.g., Gorgolewski et al.
\protect\hyperlink{ref-gorgolewski2016}{2016}), standardizing data
structures may increase research quality by clarifying concepts that are
understood or used differently by different people. When those standards
are public, they contribute to make science more open, transparent and
reproducible. Finally, the use of standards can guide the development of
various software tools that are specifically designed to take advantage
of those standards.

There are a few examples that demonstrate how sometimes even simple data
organization principles can lead to the development of an elegant and
efficient software ecosystem that greatly facilitates the analysis of
data. In the R community, for example, the notion of ``tidy'' data
(e.g., ``tidy data''; Wickham \protect\hyperlink{ref-wickham2014}{2014})
has led and contributed to the development of the suite of tools known
as the ``tidyverse'' (Wickham et al.
\protect\hyperlink{ref-wickham2019}{2019}) which has had a massive
impact on data science. Similarly, in the neuroimaging community, the
BIDS' way of organizing imaging data has had profound positive effects
for the field as whole, facilitating the sharing and reuse of imaging
data but also leading to the development of software tools to check for
example the integrity of data but also efficient and standardized data
analysis pipelines (e.g., \url{https://fmriprep.org/}; Esteban et al.
\protect\hyperlink{ref-esteban2019}{2019}). What these examples show is
that the development of standards for structuring data can lead to the
development of tools and data analysis standards that greatly benefit
the field. It is our hope that by contributing to standardizing
behavioral data, equally impressive progress can be achieved in
behavioral sciences.

In this document, we focused only on a few key concepts; other ideas are
presented in greater detail in the projects' website
(\url{https://behaverse.github.io/data-model/}) which holds an updated
version of the behaverse data model. Many questions remain unanswered,
various aspects of behavioral data to be explored and numerous decisions
to be taken. Ultimately, the value of this or any other data model will
require demonstrating that it can indeed represent rich behavioral data
across a variety of settings in a consistent way and that it offers
concrete benefits to the researchers using those standards.

\hypertarget{conclusion-1}{%
\section{Conclusion}\label{conclusion-1}}

Behavioral data is fundamental in cognitive sciences and there is
clearly a need for standards to organize such data so it can be
efficiently analyzed, shared and reused. Here we emphasized several key
issues and presented constructs we believe are essential for structuring
behavioral data and which currently seem to be used inconsistently.

Much remains to be discussed. To keep this document short and decrease
the likelihood of its content becoming obsolete as our standards evolve,
we decided to focus here only on key points and refer the reader to the
online documentation of the behaverse data model (see
\url{https://behaverse.github.io/data-model/}).

\hypertarget{acknowledgements}{%
\section{Acknowledgements}\label{acknowledgements}}

This research was supported by the Luxembourg National Research Fund:
ATTRACT/2016/ID/11242114/DIGILEARN and INTER
Mobility/2017-2/ID/11765868/ULALA grants.

\hypertarget{references}{%
\section*{References}\label{references}}
\addcontentsline{toc}{section}{References}

\hypertarget{refs}{}
\leavevmode\hypertarget{ref-braver2001}{}%
Braver, T. S., D. M. Barch, B. A. Keys, C. S. Carter, J. D. Cohen, J. A.
Kaye, J. S. Janowsky, et al. 2001. ``Context Processing in Older Adults:
Evidence for a Theory Relating Cognitive Control to Neurobiology in
Healthy Aging.'' \emph{Journal of Experimental Psychology. General} 130
(4): 746--63.

\leavevmode\hypertarget{ref-esteban2019}{}%
Esteban, Oscar, Christopher J. Markiewicz, Ross W. Blair, Craig A.
Moodie, A. Ilkay Isik, Asier Erramuzpe, James D. Kent, et al. 2019.
``fMRIPrep: A Robust Preprocessing Pipeline for Functional MRI.''
\emph{Nature Methods} 16 (1): 111--16.
\url{https://doi.org/10.1038/s41592-018-0235-4}.

\leavevmode\hypertarget{ref-gorgolewski2016}{}%
Gorgolewski, Krzysztof J., Tibor Auer, Vince D. Calhoun, R. Cameron
Craddock, Samir Das, Eugene P. Duff, Guillaume Flandin, et al. 2016.
``The Brain Imaging Data Structure, a Format for Organizing and
Describing Outputs of Neuroimaging Experiments.'' \emph{Scientific Data}
3 (1, 1): 160044. \url{https://doi.org/10.1038/sdata.2016.44}.

\leavevmode\hypertarget{ref-martin2009}{}%
Martin, Robert C., ed. 2009. \emph{Clean Code: A Handbook of Agile
Software Craftsmanship}. Upper Saddle River, NJ: Prentice Hall.

\leavevmode\hypertarget{ref-poldrack2011}{}%
Poldrack, Russell A., Aniket Kittur, Donald Kalar, Eric Miller,
Christian Seppa, Yolanda Gil, D. Stott Parker, Fred W. Sabb, and Robert
M. Bilder. 2011. ``The Cognitive Atlas: Toward a Knowledge Foundation
for Cognitive Neuroscience.'' \emph{Frontiers in Neuroinformatics} 5.
\url{https://doi.org/10.3389/fninf.2011.00017}.

\leavevmode\hypertarget{ref-robertson1997}{}%
Robertson, Ian H, Tom Manly, Jackie Andrade, Bart T Baddeley, and Jenny
Yiend. 1997. ```Oops!': Performance Correlates of Everyday Attentional
Failures in Traumatic Brain Injured and Normal Subjects.''
\emph{Neuropsychologia} 35 (6): 747--58.
\url{https://doi.org/10.1016/S0028-3932(97)00015-8}.

\leavevmode\hypertarget{ref-teeters2015}{}%
Teeters, Jeffery L., Keith Godfrey, Rob Young, Chinh Dang, Claudia
Friedsam, Barry Wark, Hiroki Asari, et al. 2015. ``Neurodata Without
Borders: Creating a Common Data Format for Neurophysiology.''
\emph{Neuron} 88 (4): 629--34.
\url{https://doi.org/10.1016/j.neuron.2015.10.025}.

\leavevmode\hypertarget{ref-wickham2014}{}%
Wickham, Hadley. 2014. ``Tidy Data.'' \emph{Journal of Statistical
Software} 59 (1, 1): 1--23. \url{https://doi.org/10.18637/jss.v059.i10}.

\leavevmode\hypertarget{ref-wickham2019}{}%
Wickham, Hadley, Mara Averick, Jennifer Bryan, Winston Chang, Lucy
D'Agostino McGowan, Romain François, Garrett Grolemund, et al. 2019.
``Welcome to the Tidyverse.'' \emph{Journal of Open Source Software} 4
(43): 1686. \url{https://doi.org/10.21105/joss.01686}.

\end{document}